# Transfer Potentials shape and equilibrate Monetary Systems

*Robert Fischer*[*] *and Dieter Braun*[+]

*Amriswilerstr. 108, CH-8590 Romanshorn, Switzerland*

[+]*Center for Studies in Physics and Biology*
*Rockefeller University, New York, USA*
*e-mail: mail@dieterb.de*

We analyze a monetary system of random money transfer on the basis of double entry bookkeeping. Without boundary conditions, we do not reach a price equilibrium and violate text-book formulas of economist's quantity theory (MV=PQ). To match the resulting quantity of money with the model assumption of a constant price, we have to impose boundary conditions. They either restrict specific transfers globally or impose transfers locally. Both connect through a general framework of transfer potentials. We show that either restricted or imposed transfers can shape gaussian, tent-shape exponential, boltzmann-exponential, pareto or periodic equilibrium distributions. We derive the master equation and find its general time dependent approximate solution. An equivalent of quantity theory for random money transfer under the boundary conditions of transfer potentials is given.

Recently, economic models of money transfer were connected to statistical mechanics [1]-[7]. Under the boundary condition of a constant number of assets, the description of wealth was related to the statistical mechanics of physics such as Boltzmann-Gibbs distributions [1][2][4] or distributions with pareto tails [2]-[4]. Evidence for this approach has been found in income data [5][6]. We further explore this statistical approach [7] to the monetary part of the economy. In contrast to previous work, we base our analysis on the more general laws of double entry bookkeeping [8]-[12]. Assets are now balanced by liabilities and the distribution of monetary wealth has an asset and a liability side. Under a most simple model of random money transfer, we confirm the Boltzmann-Gibbs distribution [1] when the liability per agent is limited. However, when we use other boundary conditions, we find gaussian, tent-shape exponential or pareto distributions although the mode of transfer has not changed.

Interestingly, random money transfer without boundary conditions contradicts economist's quantity theory. Quantity theory was the basis for monetary policies for a long time and relates the quantity of money M and the money velocity V with the price level P [9][10][18]. In quantity theory, when quantity M or velocity V rises, the price level P rises accordingly (see eq.(5)). Yet random money transfer shows a different behavior: the quantity M increases without bounds although the price level and all other parameters of quantity theory are held fix. We therefore argue that the choice of boundary condition is a major factor in determining the final distribution. Probably for many monetary systems, the dynamics of the agents only play a minor role. This would justify the approach to use random transfer schemes to model the monetary part of the economy.

We develop a potential-based approach which allows to model the boundary conditions of limited or imposed transfers. The boundary conditions are thus given by the potential. We find the master equation and are able to give an analytical time dependent approximation which converges for large times to the correct distribution. From this solution, we find that the transfer potential directly determines the final distribution. The wealth distributions can be shaped at will with a suitable transfer potential. This potential approach to random money transfer is very similar in structure to the quantum mechanics of the Schrödinger equation.

## *1. Double entry bookkeeping of money transfer*

In double entry bookkeeping, transfer of money is intrinsically linked to the creation of debt [11][14][15][16]. A monetary transfer from agent A to agent C triggers one of four different cases registered with four different bookkeeping records. The four cases are chosen based on the initial conditions of asset and liability of A and C (Fig. 1a):

1. Transfer by creation if A has liabilities and C has assets. A will increase its liabilities and C will increase its assets.
2. Transfer of asset if both A and C have assets. A will decrease and C will increase their assets.
3. Transfer of liability if both A and C have liabilities. A will increase and C will decrease their liabilities.
4. Transfer by annihilation is applied if A has assets and C has liabilities. A will decreases its assets and C decreases its liabilities.

These four cases mean that the monetary transfer in double entry bookkeeping is directly linked to creation and reduction of the quantity of money. Instead of counting assets $a_i$ and liabilities $l_i$ of agent i, we can implement all four cases numerically by using a single wealth variable

$$p_i = a_i - l_i \tag{1}$$

In a transfer, we decrease the wealth $p_A$ of A and increase the wealth $p_C$ of C. The same four cases still apply in the money transfer, but they are now hidden behind the sign arithmetic of summation and subtraction of $p_i$.

We have translated the bookkeeping records to Feynman-graphs (Fig. 1a) using a recently established connection from bookkeeping to the bouncing of particles called bookkeeping mechanics [14]-[16]. In this mechanical picture, the statistical mechanics of monetary transfer becomes the statistical mechanics of a gas. Differing from an ideal gas, the monetary gas does not implement conservation of energy. Some transfer collisions will lead to particle creation or annihilation.

The quantity of money in bookkeeping can be generalized from economics by counting the absolute value $M_{ic} = |p_{ic}|$ of the number of currency units $p_{ic}$ of agent i and currency c. This definition supersedes the economical definition which counts only certain bank deposits. We can infer the change in M induced by a transfer. We have to split the transferred currency units $\Delta p_{ic}$ into a change from increase $\Delta p_{ic}^{a+} \geq 0$ or decrease $\Delta p_{ic}^{a-} \leq 0$ of assets and into an increase $\Delta p_{ic}^{l+} \leq 0$ or decrease $\Delta p_{ic}^{l-} \geq 0$ of liabilities ($\Delta p_{ic} = \Delta p_{ic}^{a+} + \Delta p_{ic}^{l-} + \Delta p_{ic}^{a-} + \Delta p_{ic}^{l+}$):

$$M_{ic} = |p_{ic}| \qquad \Delta M_{ic} = \Delta p_{ic}^{a+} - \Delta p_{ic}^{l-} + \Delta p_{ic}^{a-} - \Delta p_{ic}^{l+} \tag{2}$$

Bookkeeping mechanics would motivate a quadratic measure of the quantity analogous to the energy in mechanics by choosing $E_{ic} = p_{ic}^2/2$. The change in energy can be inferred without splitting $\Delta p_{ic}$:

$$E_{ic} = p_{ic}^2/2 \qquad \Delta E_{ic} = (\Delta p_{ic})^2/2 + p_{ic}\Delta p_{ic} \tag{3}$$

The novelty in both quantity definitions $M_{ic}$ and $E_{ic}$ is that we can identify the agent i who changed the quantity of currency c. Quantity becomes a microscopic variable. In the random model discussions below, we will calculate M and E from the distribution of monetary wealth n(p,t):

$$M = \langle M_i \rangle = \int_{-\infty}^{\infty} |p| n \, dp \qquad E = \langle E_i \rangle = \int_{-\infty}^{\infty} \frac{p^2 n}{2} dp \qquad (4)$$

By doing so, we exclude the assets and liabilities of the bank and count only their mirrored part in the bookkeepings of the agents. The bank is not considered to be counted as an agent in the following models.

## 2. Economy of random money transfer: Contradiction to quantity theory

The random transfer of the model is simple: Each of the N agents indexed by i have no initial currency units: $p_i(t=0)=0$. They choose transfer partners randomly at each time step t, yielding N transfer pairs. Each pair transfers $\Delta p$ currency units. Such a simplified model is motivated from the finding that details of the monetary exchange models do not appear to be essential [1]. By choosing a random transfer scheme, we implement the most simple money transfer economy which does not follow the dogma of money moving in circles. For example random money transfer implements N farmers without surplus, but a fluctuating harvest of $p_0-\Delta p$, $p_0$ or $p_0+\Delta p$ given in monetary units. Each year, they balance their luck by exchanging harvest against money.

Random money transfer leads to an expansive diffusion process with a diffusion constant given by

$$D = (\Delta p)^2 / \Delta t \qquad (5)$$

As expected, the distribution of monetary units n(p,t) is gaussian and the total quantity of money M follows a square root time evolution:

$$\dot{n} - D\Delta n = 0 \qquad n = \frac{N}{\sqrt{4\pi Dt}} \exp{-\frac{p^2}{4Dt}} \qquad M = 2N\sqrt{Dt/\pi} \qquad E = NDt \qquad (6)$$

The analytical results have been checked against a numerical simulation [17]. We show distributions for N=2000 for $\Delta p=1$ at $t=100\Delta t$, $200\Delta t$ and $300\Delta t$ together with the time evolution of the quantity of money per agent M/N in Fig. 1b.

We count the frequency of the applied bookkeeping records. We find only creations or annihilations for N=2, transfers of assets and liabilities appear only for N>2. Creations occur more often than annihilations, accounting for the same increase in M, independent of N. This is due to the asymmetry of the initial conditions. For a transfer between two agents with p=0 each, only a creation and no annihilation can be performed.

This basic model of random money transfer contradicts text-book versions of quantity theory. With M the money 'supply', V the money velocity, P the price 'level' and Q the 'real' gross national product (GNP), quantity theory [9][10][18] states

$$MV = PQ \qquad (7)$$

The random money transfer yields a fixed $V=\Delta p/\Delta t$, a fixed price level $P=\Delta p$ and a fixed GNP given by the number of agents N multiplied by the exchanged prices over the time span $Q=N\Delta p/\Delta t$. We thus expect a constant $M=N\Delta p$. Yet, we find an increase of M without bounds. Quantity theory is therefore not applicable to random money transfer which creates of asset-liability pairs from loans. Such important restrictions were known to the inventors of quantity theory, but subsequently dropped (see discussion in [18], p. 40ff).

In the case when agents transfer through a bank B, the bookkeeping splits into two independent currencies. One is used for the assets of the agents (black deposit currency) and the other for the liabilities (white loan currency) of the agents as shown in the bookkeeping records (Fig. 1c) and discussed thoroughly elsewhere [16]. In such a bicurrency system, a transfer between agents involves two prices: a deposit price if paid by deposits and a loan price if paid from a loan. As an example, we choose a transfer of $\Delta p_D=1$ deposit currency units and $\Delta p_L=2$ loan currency units, establishing an exchange rate of 1:2. In this case, the liability side of distribution expands faster with twice the quantity of money (Fig. 1d). Note that the discontinuity at p=0 in the distribution is no artefact, since we apply the four bank bookkeeping records of the bicurrency system (Fig. 1c). It means we cannot use anymore the compactification of (eq. 1), but have to account assets and liabilities independently. Although it would be instructive (and easily possible) to use a bicurrency system in the following, we will restrict ourselves again to one currency which describes the bank bookkeeping practice of today [16].

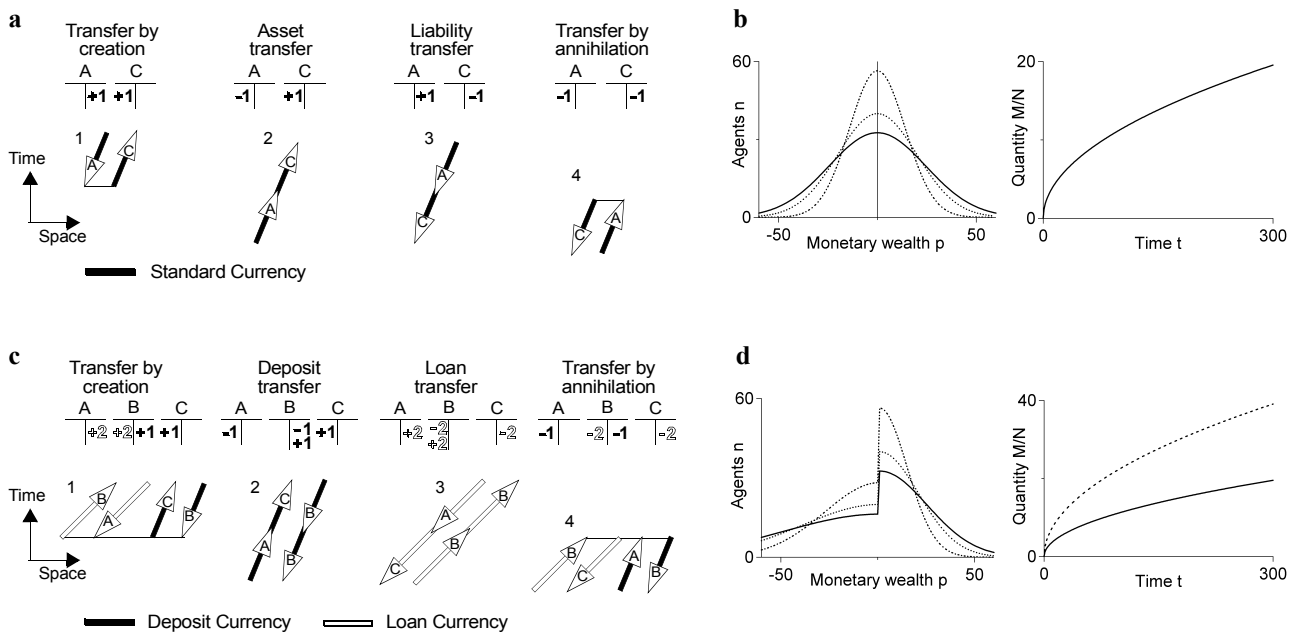

Fig. 1: Bookkeeping of money transfer and random money transfer. (a) Money transfer from agents A to C without a bank yields four possibilities depending on the initial stock of asset and liability of A and C. (b) Random money transfer results in an expanding gaussian distribution of monetary wealth. The quantity of money increases without bounds. (c) The four cases of money transfer from agents A to C through a bank B establishes a bicurrency system of deposit currency and loan currency [16]. (d) For an exchange rate of 1:2 between deposit currency and loan currency, we find a step in the distribution. The quantity of both currencies differ by a factor of 2.

## *3. Forbidden transfers equilibrate the random economy globally*

We have seen so far, that a small economy with no surplus, but with a random choice of transfer partners will increase the quantity of money without bounds. In the following we will discuss, how boundary conditions can equilibrate the system. In most cases, a constant quantity level M will be reached. We will try to match the results with quantity theory

We can impose an equilibrium by forbidding transfers which would increase the quadratic measure of the quantity of money E over a target $E_0$. This is a global, non-local approach. Before each transfer, it has to be tested whether the total E will be increased over the limit $E_0$. Which such a transfer restriction, the distribution converges to a gaussian profile ($E_0=25\pi N$; Fig. 2a: distribution at $t=300\Delta t$, Fig. 2b solid line: quantity M over time):

$$ n = \frac{N}{\sqrt{4\pi E_0/N}} \exp-\frac{Np^2}{4E_0} \qquad M = 2\sqrt{\frac{E_0 N}{\pi}} \qquad (8) $$

Another way to equilibrate the random monetary system is to forbid all transfers which would increase the total quantity of money over a target quantity $M_0$. Now the random economy converges to a tent-shaped exponential distribution ($M_0=10N$; Fig. 2d: distribution at $t=300\Delta t$, Fig. 2e solid line: quantity M over time):

$$ n = \frac{N^2}{2M_0} \exp-\frac{N|p|}{M_0} \qquad M = M_0 \qquad (9) $$

For the last two cases of transfer restrictions, we could find the result of quantity theory if we would explicitly assume a scaling of $M = N\Delta p$, meaning that the limit M or E would somehow know the price level $\Delta p$ of the system.

Typically in model systems, a liability limit is applied to simulations of monetary systems [1]-[7]. We simulate also this case and forbid all transfers which would increase the liabilities of an agent above a limit L. This loan restriction leads to an exponential Boltzmann distribution n(p) in accordance with the results in the literature [1]-[7] (L=5e; Fig. 2g: distribution at t=300Δt, Fig. 2h solid line: quantity M over time):

$$n = \frac{N}{L}\exp-\frac{p+L}{L} \qquad M = 2NLe^{-1} \qquad (10)$$

The quantity theory (eq. 7) would be fulfilled, if the liability limit would explicitly scale with the price level Δp as L=eΔp/2, yielding a quite low liability limit of L=1.36 for Δp=1. Note that this boundary condition was thoroughly discussed by Dragulescu and Yakovenko [1].

We can impose a pareto distribution by defining a logarithmic money quantity $G_{ic} = \ln|p_{ic}|$ and forbidding all transfers which would increase the total G above a limit $G_0$. In this case the distribution tries to converge to a pareto distribution ($G_0$=0.8N; Fig. 2k: distribution at t=300Δt, Fig. 2l solid line: quantity M over time):

$$n = k|p|^{-1/D} \qquad M \to \infty \qquad (11)$$

The quantity of money M is slowly diverging. Since the pareto distribution cannot be normalized [20], we had to fit k=215 to match the grain and timing of our numerical implementation. Furthermore the simulation assumed G=0 for p=0. Still, the pareto distribution in its hyperbolic form is found. The tails of the distribution are not fully filled by the diffusion process at t=300Δt. A change in price level Δp affects D and therefore the functional shape of the distribution.

We have to note that nonlinear quantity definitions such as E or G work only with properly indexed agents i. Splitting the wealth of one agent into the wealth of subagents would change the total $E_0$ or $G_0$. This problem is solvable: all bookkeeping must be directly connected to real agents.

## *4. Imposed transfers equilibrate the random economy locally: Transfer potentials*

Until now, the boundary conditions forbid specific transfers. A second class of boundary conditions allow all transfers, but act on the stock of assets and liabilities. This gives us a local approach. We do not have to measure the total quantity of money before each transfer. We will show how locally imposed transfers lead to the identical equilibrium of the global method of forbidden transfers as discussed before.

For example, we reach an equilibrium by imposing a negative interest rate r to both assets and liabilities. The unit of interest is % per infinitesimal time interval δt. Within macroscopic time Δt, the interest application changes the assets and liabilities by a factor of $f = \exp(r\Delta t) = \tilde{r}\Delta t + 1$ which defines a finite-time interest rate $\tilde{r}$ given in % per finite time interval Δt. Random money transfer now converges to a time independent gaussian profile (r=-1/(50πδt)=-0.64%/δt, Fig. 2c, right: distribution at t=300Δt):

$$n = N\sqrt{\frac{-r}{2\pi D}}\exp\left[\frac{rp^2}{2D}\right] \qquad M = 2N\sqrt{\frac{D}{-2\pi r}} \qquad E = \frac{ND}{-r} \qquad (12)$$

For the first time, we recover the price scaling of quantity theory: $M \propto N\sqrt{D} \propto N\Delta p$. We find the exact identity of quantity theory (eq. 7) for a rather high negative interest rate of $r = -2/\pi\delta t = -64\%/\delta t$. Negative interest rates realize the implicit price scaling of quantity theory.

This example gives us a working model. We might be able to reach the same distribution by forbidding transfers or by imposing transfers. The imposed transfers are local and are not a function of the price level Δp or the transfer velocity Δp/Δt. Nevertheless the final distribution does depend on D as expected from a monetary system in which the wealth distribution reflects the price level. We will show in the following that both approaches to a monetary equilibrium can be connected by reinterpreting the quantity of money definition (M,E or G) as transfer potential U. We will be able to derive and to approximate the general master equation of potential restricted random money transfer.

We introduce a transfer potential U(p). It models all additional imposed external transfers F given by the transferred amount Δp per time interval Δt. We define F by the negative derivative of a potential U(p) as usual in physics:

$$F(p) = \frac{\Delta p}{\Delta t} = -\nabla U(p) \qquad (13)$$

One should think of F as an interest rate progression table or a legal system to impose tax inputs and outputs depending only on the wealth p of the agents. Typically, we would infer U(p) from integrating the imposed transfers F(p). Adding the monetary flow nF(p) from the transfers F to Fick's first law, we find:

$$j = -D\nabla n - n\nabla U(p) \qquad (14)$$

The potential U has the same units as the diffusion constant D. The approach is quite similar to the treatment of thermophoresis with U being the temperature [19]. Within the logics of bookkeeping mechanics [14][15][16], the units for D and U would be power and for F force. By inserting Fick's law as usual into the continuity equation, we derive the master equation of random money transfer under a transfer potential:

$$\dot{n} - D\Delta n - \nabla[n\nabla U(p)] = 0 \qquad (15)$$

We can give a general time dependent approximation when all agents start with p=0 at t=0:

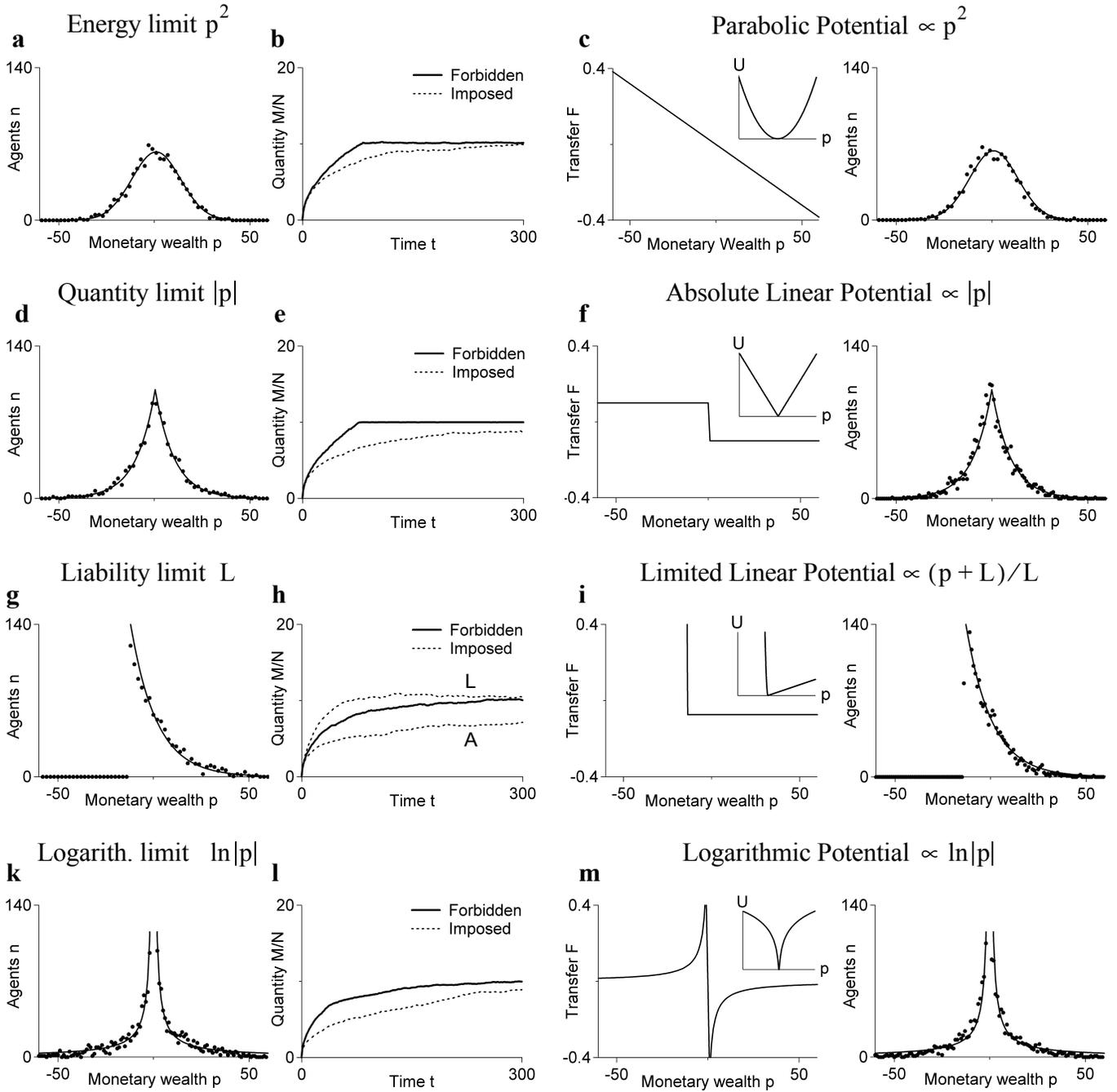

Fig. 2: Different boundary conditions equilibrate random money transfer to different distributions. (a-c) We forbid transfers which increase the energy of money over $E_0$ and find a gaussian distribution, also found from imposed transfers using a parabolic potential. (d-f) We forbid transfers which increase the quantity of money over $M_0$ and find a tent-shaped exponential distribution, also found from imposed transfers using a absolute-linear potential. (g-i) We forbid transfers which increase a non-bank liability over a limit L and find a Gibbs-Boltzmann exponential distribution, also found from imposed transfers using a limited-linear potential. (k-m) We forbid transfers which increase a logarithmic quantity definition over $G_0$ and found a pareto distribution, also found from imposed transfers using a logarithmic potential.

$$n(p, t) \cong N \frac{\exp-\frac{p^2}{4Dt}}{\sqrt{4\pi Dt}} \frac{\exp-\frac{U(p)}{D}}{\int_{-\infty}^{\infty} \exp[-U(p)/D]dp} \quad (16)$$

The approximation is valid, when the term $p\nabla U/Dt$ becomes negligible. For large times the distribution is therefore only determined by the potential U and the diffusion constant D:

$$n(p) \propto \exp\left(-\frac{U(p)}{D}\right) \quad (17)$$

The diffusion constant D merely defines the units in which the potential U is given. Changes in the price level will change D and only rescale the distribution with the exponent $-1/D$ according to (eq. 16) and (eq. 17).

The potential U(p) might impose an asymmetry onto the final distribution (eq. 17). The quantity of money M could differ for the total assets $M_A$ and for the total liabilities $M_L$. This means that an external transfer $p_{ext}$ was extracted (-) or injected (+) from the system between the times $t_1$ and $t_2$:

$$p_{ext}(t_1, t_2) = \int_{-\infty}^{\infty} p[n(t_2) - n(t_1)]dp \quad (18)$$

Logically, when $p_{ext} \neq 0$, an additional agent out of the random transfer has to be defined to account for the missing assets and liabilities. One would expect that such an additional agent is a bank or the state. For example, we could use $p_{ext}$ to model deficit spending. In the following however, we discuss only potentials U which impose no total external transfer $p_{ext}$.

In the first part of the paper, we have shown four boundary condition scenarios which restricted random money transfer (eq. 8)-(eq. 11). Now we achieve the same final distributions without restricting any transfers, but by imposing additional transfers (eq. 13)-(eq. 17). The transfer potential biases the random money transfer to comply with the boundary conditions:

  (i)   Parabolic potentials implement an energy limit and a gaussian distribution (Fig. 2c)
  (ii)  Absolute-linear potentials realize a quantity limit and a tent-shape exponential distribution (Fig. 2f)
  (iii) Limited-linear potentials impose a liability limit and a boltzmann-exponential distribution (Fig. 2i)
  (iv)  Logarithmic potentials show a logarithmic limit and a pareto distribution (Fig. 2m).

In each case we did a numerical calculation for N=2000, Δp=1, D=1 until t=300Δt. We show the quantity M/N over time (dashed line in Fig. 2b,e,h,l) together with the used transfer F and potentials U and the final distribution in Fig. 2c,f,i,m. The final distribution is fitted with the analytical limit solution (eq. 17) of the master equation (eq. 15), given by a solid line in (Fig. 2c,f,i,m). The imposed potentials U and used transfers F are as follows:

$$
\begin{array}{ll}
\text{(i)} & U = -\dfrac{rp^2}{2} \qquad\qquad F = rp \\[6pt]
\text{(ii)} & U = F_0|p| \qquad\qquad F = -\mathrm{sgn}(p)F_0 \\[6pt]
\text{(iii)} & U = \begin{cases} \infty & ,p<-L \\ \dfrac{D(p+L)}{L} & ,p>-L \end{cases} \qquad F = \begin{cases} {}'\infty' & ,p<-L \\ -\dfrac{D}{L} & ,p>-L \end{cases} \\[6pt]
\text{(iv)} & U = \ln|p| \qquad\qquad F = -\dfrac{1}{p}
\end{array}
\qquad (19)
$$

They yield the following master equations and their limit distribution n(p) for time $t \to \infty$:

$$
\begin{array}{lll}
\text{(i)} & \dot{n} - D\Delta n + 2r[n + p\nabla n] = 0 & n(p) = N\sqrt{\dfrac{-r}{2\pi D}}\exp\!\left(\dfrac{rp^2}{2D}\right) \\[8pt]
\text{(ii)} & \dot{n} - D\Delta n - \mathrm{sgn}(p)F_0 \nabla n = 0 & n(p) = N\dfrac{F_0}{2D}\exp\!\left(-\dfrac{F_0|p|}{D}\right) \\[8pt]
\text{(iii)} & \dot{n} - D\Delta n - \dfrac{D\nabla n}{L} = 0 & n(p) = \dfrac{N}{L}\exp\!\left(-\dfrac{p+L}{L}\right) \\[8pt]
\text{(iv)} & \dot{n} - D\Delta n - \nabla\!\left[\dfrac{n}{|p|}\right] = 0 & n(p) = k|p|^{-1/D}
\end{array}
\qquad (20)
$$

We compare these results of imposed transfers with the results of forbidden transfers (eq. 8)-(eq. 11):

(i) The parabolic potential matches (eq. 8), with $E_0 = -ND/r$ and $\langle U \rangle = ND/2$.

(ii) The absolute linear potential matches the discussion of (eq. 9) with $F_0 = ND/M_0$ and $\langle U \rangle = ND$.

(iii) The limited-linear case directly matches (eq. 10) and we find again $\langle U \rangle = ND$. The limited-linear case was simulated with a steep parabolical potential at p=-L to allow numerical treatment. The liabilities (Fig. 2h, L) converge much faster to the equilibrium than the assets (A), which lag behind. Both eventually reach the same level with $p_{ext}=0$ for $t \to \infty$ (eq. 18).

(iv) The logarithmic potential matches the results of imposed transfers (eq. 11). We have to apply upper and lower boundaries to F for small |p| to prevent divergence.

The potential which imposed the distribution is identical to the quantity definition which forbid the transfers: (i) U=E; (ii) U=M and (iv) U=G. This is no coincidence. When a transfer is forbidden it only means that without restrictions the transfer would have been done, but the imposed transfers immediately reversed it. We can generalize this notion under the assumption that the monetary flow j, given in (eq. 14), is zero in the steady state of the equilibrium ($\dot{n} = 0$). We derive by partial integration:

$$\langle U \rangle = ND - D \int_{-\infty}^{\infty} \frac{U \Delta U}{(\nabla U)^2} n \, dp \qquad \text{for} \qquad \lim_{p \to \pm \infty} \frac{Un}{\nabla U} = 0 \qquad (21)$$

The transfer restriction given by (eq. 21) on the total quantity measure $\langle U \rangle$ is identical to imposing a transfer potential U in a random transfer environment. Take for example the parabolic potential of interest rates r given by $U=-rp^2/2$. As we change the interest rate r, we also change the quadratic measure of the quantity $U_{ic} = -rp_{ic}^2/2$. Therefore, since the interest rate changes our measure of the quantity of money $U_{ic}$, we still reach a constant steady state at $\langle U \rangle = ND/2$ given by (eq. 21).

We call the relationship (eq. 21) the quantity theory of random money transfer under boundary conditions. It connects the measurement functional of monetary quantity U with the number of agents N and the market diffusion parameter D. The units of the equation (NΔp²/Δt) are identical to the classical quantity equation (eq. 7).

In a way, (eq. 21) resembles the equipartition theorem of statistical mechanics with U the energy and D the temperature. With D as temperature, we can reinterpret the equilibrium distribution of (eq. 17) as boltzmann distribution of statistical mechanics: the structure of statistical thermodynamics appears. We use it to define the sum of states $Z(t) \cong \int_{-\infty}^{\infty} \exp[-\tilde{U}/D] \, dp$ over time with energy $\tilde{U}(p, t) = p^2/4t + U(p)$ to find the free energy $F = -DN \ln Z$ and recover the distribution of (eq. 16) from $n = \partial F / \partial \tilde{U}$. The entropy $S = -\int_{-\infty}^{\infty} n \ln(n) \, dp$ can be derived from $S = -\partial F / \partial D$ and the time dependent version of (eq. 21) from $\langle U \rangle = ND^2/Z \cdot \partial Z / \partial D$.

By imposing transfers instead of forbidding transfers, no difficult measurement has to be undertaken for of the price level Δp, the diffusion constant D and the number of agents N. The monetary system will converge to a distribution depending on D and N, although the regulatory transfers $F = -\nabla U$ do not depend on the market parameters Δp, N and D. This is true for the absolute linear case ($M \neq M(\Delta p, N, D)$) and the parabolic case of interest rates ($E \neq E(\Delta p, N, D)$), but for example not for a liability limit. Here the potential U depends on the diffusion constant D to converge without external transfers $p_{ext}=0$ (eq. 18).

The second advantage of imposed transfers from a transfer potential is the local nature of the equilibrium regulation. Only the local rule (eq. 13) has to be implemented, the total quantities M, E or G do not have to be measured. This is a very important practical advantage of imposed transfers.

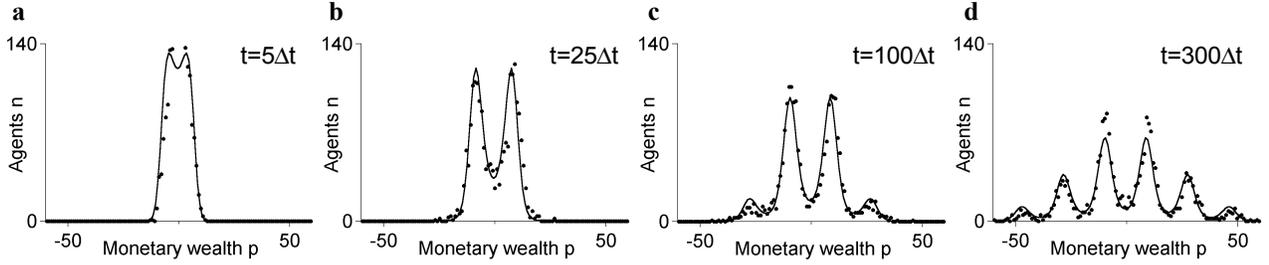

Fig. 3: Random money transfer under a sinusoidal transfer potential. The general time dependent approximation [(eq. 16), solid line] describes well the histogram of random money transfer shown as dots. The imposed transfers yield distribution maxima at the potential minima according to (eq. 17).

In the real world, the transfer potentials $U(p_i,i)$ among the agents i are defined by the bias of the subjects to deviate from a random money transfer, for example induced by the input and output characteristics of banks and the government. We applied here the approximation that the transfer potential $U(p_i)$ only depends on the wealth p of the agent i. Secondly we assume that the economy can be described as agents which are forced by an outside environment to do random transfers of money. On a random money transfer assumption, we would be able to calculate from tax laws and interest rate characteristics the final wealth distribution n(p) according to (eq. 17).

Towards the end, we want to demonstrate that the potential theory can shape all kinds of wealth distributions. It should also show that the general time dependent approximation of (eq. 16) is quite precise. Assume that we impose a sinusoidal transfer potential U:

$$U = \cos(p/3) \qquad F = \frac{1}{3}\sin(p/3) \tag{22}$$

We simulate this monetary system with the usual parameters N=2000, D=1 and show the simulated distribution together with the theoretical approximation from (eq. 16) at times t=5Δt, 25Δt, 100Δt and 300Δt in Fig. 3a,b,c,d. The imposed transfers F stamp a grating into the wealth distribution as the diffusion spreads without converging to an equilibrium. The approximation of (eq. 16) describes the simulation results over time with high precision.

Any one familiar with solutions of the Schrödinger equation will spot similarities between potential restricted random money transfer and quantum mechanics. The 'wave packet' diffuses without a potential. A gaussian distribution is found for a parabolic potential. The time dependent solution yields an exponential prefactor. Energy is allowed to fluctuate. This is probably just a coincidence due to the diffusive similarities of (eq. 15) and the Schrödinger equation. Also note that we are operating here in momentum space according to the analogy of bookkeeping mechanics. Nevertheless the simi-larities should be very helpful in approaching random money transfer from physics. We have chosen variable letters from physics to enhance this transition. The physical analog is a gas [14] where particles bounce, create and annihilate under random momentum transfer and momentum conservation. Energy conservation is not achieved unless a suitable transfer potential is imposed. The potential models a viscous friction by a medium to allow energy conservation.

## *5. Conclusions*

We have discussed an economy of random monetary transfer between agents. We based our analysis on explicit bookkeeping records of assets and liabilities, shown as Feynman-graphs using the momentum analogy of bookkeeping mechanics. Interestingly, we cannot reach a price equilibrium without restrictive boundary conditions. The boundary conditions either forbid certain transfers or impose transfers. We can switch between both pictures analytically by identifying the monetary quantity definition as transfer potential. We find a variety of distributions by using different potentials. With the approach of imposing a potential, we have a powerful technique to describe distributions and their underlaying imposed transfers. The master equation of random transfer (eq. 15) converges to the general distribution (eq. 17) and allows to discuss analytical results for a large variety of biased random money transfer scenarios. Since the approach is so near to the mathematics of quantum mechanics, we expect many more insights and cross-transfers from physics. Moreover, the given potential approach can be applied to other quantities which are exchanged or regulated in an economy.

We have found that any application of quantity theory has to operate under imposed boundary conditions. Central banks have changed their definition of the quantity of money several times. The FED and the swiss central bank eventually gave up to define it. Our findings suggest that the boundary conditions are essential and their change over time has to be considered. When central banks want to regulate monetary value over time, they may profit from the shown general potential theory of monetary systems.

The authors would like to thank Koichiro Matsuno, Victor Yakovenko and Benjamin Franksen for comments and Klaus-Peter Karmann, Veronica Egger and Noel Goddard for reading earlier drafts of the manuscript.